\documentclass{article}
\usepackage{spconf,amsmath,graphicx,amssymb}
\usepackage{enumitem}
\usepackage{hyperref}
\setlist{nosep, leftmargin=14pt}

\usepackage{mwe} 
\usepackage{times}

\usepackage{caption}

\newcommand{\smallCross}{\textsuperscript{\textdagger}}

\title{3D Anatomical Structure-Guided Deep Learning for Accurate Diffusion Microstructure Imaging}
\name{Xinrui Ma$^{1,2}$, \ Jian Cheng$^{3,4}$, \ Wenxin Fan$^{1,2}$,  \ Ruoyou Wu$^{1,2,5}$,  \ Yongquan Ye$^{6}$,  \ \textit{Shanshan Wang}$^{1,\dagger}$ 
\thanks{\smallCross\ Corresponding author: ss.wang@siat.ac.cn}}

\address{$^{1}$Paul C. Lauterbur Research Center for Biomedical Imaging,\\ Shenzhen Institutes of Advanced Technology, Chinese Academy of Sciences, Shenzhen, Guangdong, China \\
    $^{2}$University of Chinese Academy of Sciences, Beijing, China\\
    $^{3}$State Key Laboratory of Software Development Environment, Beihang University, Beijing, China. \\
    $^{4}$Key Laboratory of  Data Science and Intelligent Computing, \\Institute of International Innovation, Beihang University, Hangzhou, Zhejiang, China\\
    $^{5}$Pengcheng Laboratory, Shenzhen, Guangdong, China\\
    $^{6}$United Imaging Healthcare, Houston, TX, USA\\
}

\begin{document}
\ninept
\maketitle
\begin{abstract}
Diffusion magnetic resonance imaging (dMRI) is a crucial non-invasive technique for exploring the microstructure of the living human brain. Traditional hand-crafted and model-based tissue microstructure reconstruction methods often require extensive diffusion gradient sampling, which can be time-consuming and limits the clinical applicability of tissue microstructure information. Recent advances in deep learning have shown promise in microstructure estimation; however, accurately estimating tissue microstructure from clinically feasible dMRI scans remains challenging without appropriate constraints. This paper introduces a novel framework that achieves high-fidelity and rapid diffusion microstructure imaging by simultaneously leveraging anatomical information from macro-level priors and mutual information across parameters. This approach enhances time efficiency while maintaining accuracy in microstructure estimation. Experimental results demonstrate that our method outperforms four state-of-the-art techniques, achieving a peak signal-to-noise ratio (PSNR) of 30.51±0.58 and a structural similarity index measure (SSIM) of 0.97±0.004 in estimating parametric maps of multiple diffusion models. Notably, our method achieves a 15× acceleration compared to the dense sampling approach, which typically utilizes 270 diffusion gradients.
\end{abstract}                
\section{Introduction}
\label{sec:intro}
Diffusion magnetic resonance imaging (dMRI) is widely applied to non-invasively estimate brain tissue microstructure by measuring the restricted diffusion of water molecules in the local microstructural environment. By carefully designing signal models that relate tissue microstructure to diffusion signals, the organization of the neuronal tissue can be inferred by fitting these models to the observed measurements. For example, diffusion tensor imaging (DTI) \cite{basser1994mr} is a classic signal model that assumes Gaussian diffusion of water molecules in voxels. Although DTI is a commonly used signal model, it is known to lack specificity \cite{pasternak2018advances}. To improve the specificity of dMRI in quantifying tissue complexity and heterogeneity, advanced diffusion models like neurite orientation dispersion and density imaging (NODDI) \cite{zhang2012noddi} have been developed. Notably, rotationally invariant scalar measures from NODDI have been widely used in neuroscientific research. However, fitting the NODDI model currently requires a large number of diffusion gradients (about 100 or more), resulting in a long acquisition time (approximately half an hour for 270 gradients) \cite{pasternak2018advances, li2024deep}. Clinical scan time for dMRI is often limited to a few minutes, typically allowing for only about 30 diffusion gradients or fewer \cite{pasternak2018advances}. Morover, dMRI scans have the problems of low signal-to-noise ratio (SNR) and reproducibility\cite{coelho2019resolving}. Recently deep learning (DL) techniques have been applied to accelerate diffusion MRI, yielding promising results\cite{wang2024knowledge}. The seminal work of Golkov et al. \cite{golkov2016q} introduced the deep learning concept to diffusion MRI and established the q-space deep learning (q-DL) framework, which has greatly benefited the subsequent work in this field. q-DL and subsequent studies have also demonstrated the promise of deep learning in using a small amount of diffusion data to predict high-quality scalar diffusion metrics from DTI \cite{tian2022sdndti, goodwin2023patch, fan2024aid} and more advanced diffusion analysis models, such as  NODDI \cite{golkov2016q,gibbons2019simultaneous,ye2019deep,xiao2024robnoddi}.

To better understand brain microstructure, integrating features from multiple diffusion models is clinically beneficial, though it increases the demands of model fitting and computational cost. To tackle these challenges, HashemizadehKolowri et al. \cite{hashemizadehkolowri2022jointly} compared three different DL approaches proposed to jointly estimate parametric maps of DTI, diffusion kurtosis imaging (DKI), NODDI, and multi-compartment spherical mean technique (SMT). Fan et al. \cite{fan2024deepmpmri} proposed DeepMpMRI, equipped with a newly designed tensor-decomposition-based regularizer, for simultaneously estimating parametric maps of DTI and NODDI.\, Li et al. \cite{li2024deep} reconstructed tissue microstructure from four brain dMRI datasets using the NODDI and SMT models, providing evidence for the clinical feasibility of DL-based tissue microstructure reconstruction.  

Integrating anatomical priors with diffusion images is essential for enhancing the directionality and accuracy of diffusion signal interpretations \cite{yendiki2022post,wu2024csr}. Chen et al. \cite{chen2023super} proposed DCS-qL, a model that incorporates multiple network branches to process patches with distinct tissue labels, thereby enhancing tissue-specific representation and diffusion signal interpretation. Huang et.al \cite{huang2023super} proposed a novel method, namely PAK-SRR, to employ rich prior anatomical knowledge (PAK) as strong guidance to improve the super resolution reconstruction of fetal brain MRI.\, Li et al. \cite{li2023diffusion} introduced DeepAnat to synthesize high-quality T1-weighted(T1w) anatomical images directly from diffusion data, enabling brain segmantation and facilitating co-registration in scenarios where T1w images are unavailable. Inspired by these pioneering works, we consider key anatomical priors based on tissue probability maps to guide microstructure estimation from multiple diffusion models.

To this end, we propose a deep learning-based framework for brain microstructure estimation that incorporates rich prior anatomical knowledge as an additional regularization term. This framework aims to simultaneously estimate parameters derived from diverse diffusion models using sparsely sampled q-space data. First, we utilize anatomical priors from three tissue probability maps obtained from T1w images. By aligning diffusion signals with anatomical boundaries, this approach effectively mitigates inaccuracies in parameter estimation that can arise from blurry tissue boundaries in the brain. Second, we have developed a highly extensible framework capable of accommodating various diffusion models. This framework employs a flexible network architecture as its backbone, enabling the leveraging of mutual information across parameters to enhance overall accuracy.
 
\section{Method}
\label{sec:format}
\subsection{Task Formulation}
By applying a set of diffusion gradients, we can acquire a vector of diffusion signals at each voxel. Each diffusion signal can be considered as a set of $W\times H\times S$ size volumes captured in the q-space. Thus the dMRI data are 4D signals of size $\mathbb{R} ^{W\times H\times S\times D } $, where $W$,$H$, $S$, $D$ refer to the width, height, slice, and gradient directions, respectively.
We aim to estimate the multiple parameters derived from various diffusion models using sparsely sampled q-space data. Given the diffusion MRI data $\mathcal{X} \in \mathbb{R} ^{W\times H \times S \times D_{Full} }$, containing the full measurements in the q-space, we can obtain the corresponding ground-truth scalar maps $\mathcal{Y}_{GT}$ by model fitting. The network $\mathcal{F} _{\theta } $ parameterized by $\theta $ is designed to learn a mapping from the given sparse sampling data  $\widetilde{\mathcal{X}} \in \mathbb{R} ^{W\times H \times S \times D_{Sparse} }$ to predicted multi-parameters $\mathcal{Y}_{Pred}$ s.t. $\mathcal{Y}_{Pred} =\mathcal{F}_{\theta } (\widetilde{\mathcal{X} } )$  $\overset{} {\rightarrow}$ $\mathcal{Y}_{GT}$. We estimate multi-parametric scalar maps by minimizing the following cost function:
\begin{equation}
\min_{\theta } L_{Data}(F_{\theta }(\widetilde{\mathcal{X}}), \mathcal{Y}_{GT})
\end{equation}
where the first term is a data fidelity term used for penalizing the difference between the network output $F_{\theta }(\widetilde{\mathcal{X}})$ and ground truth $\mathcal{Y}_{GT}$.

\subsection{Anatomical prior module}
In clinical practice, different MR imaging settings will produce various multi-contrast images, which can provide complementary information to each other. This paper introduces an anatomical prior module as the additional regularization term for preserving structural details. Specifically, we extract T1w images using the hidden Markov random field model available in the 3D/4D+ imaging library in Python (DIPY)\cite{garyfallidis2014dipy}, categorizing voxels into three types of tissues: gray matter (GM), white matter (WM), and cerebrospinal fluid (CSF). This process generates tissue probability maps, as illustrated in Fig.1. It is noteworthy that each voxel’s probability value ranges from 0 to 1, representing the likelihood that the voxel belongs to the corresponding tissue type. For example, in the white matter probability map, a voxel’s value indicates its probability of being white matter.

\subsection{Loss function}
The objective function is defined based on the network output to enforce the consistency between the predicted $\mathcal{Y} _{Pred}$ and the ground truth scalar maps $\mathcal{Y}_{GT}$ obtained from the full sampling. Different from \cite{fan2024deepmpmri}, our loss function is defined as:
\begin{equation}
\mathrm{Loss} = \left \| \mathcal{Y}_{GT} - \mathcal{Y}_{Pred} \right \|_{2}^{2}
\end{equation}
where $\left \| \cdot  \right \| _{2}^{2} $ denotes the squared L2 norm, used to compute the relative error between the predicted and ground truth values.

\begin{figure}[!t]   
  \centering
  \includegraphics[width=\linewidth]{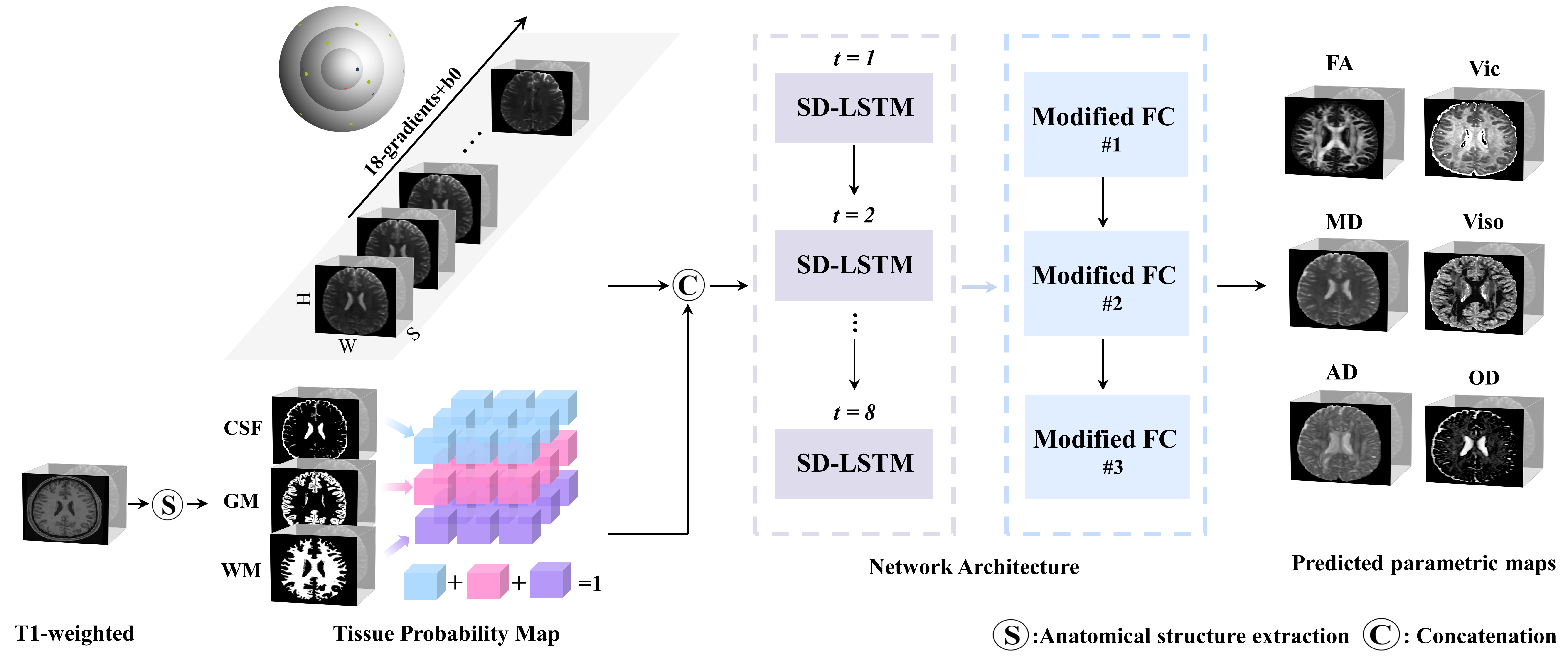}
  \caption{An overview of the proposed framework. The input consists of two components: the upper branch with six DWIs and a b0 image, and the lower branch with tissue probability maps from T1w extraction. These inputs are concatenated and processed by the network to yield DTI and NODDI microstructural parameters, including fractional anisotropy(FA), mean diffusivity(MD), axial diffusivity(AD), intracellular volume fraction(Vic), isotropic volume fraction(Viso) and orientation dispersion index(OD).}
  \label{fig1}
\end{figure}

\section{EXPERIMENTS}
\label{sec:format}
\subsection{Dataset}
This study employs preprocessed whole-brain diffusion MRI data from the publicly available Human Connectome Project (HCP) young adult dataset\cite{van2013wu}, acquired on a 3T MR scanner. A total of 111 subjects were randomly selected: 60 for training, 17 for validation, and 34 for testing. The diffusion MRI protocol consisted of three diffusion-weighted shells with b-values of 1000, 2000, and 3000 s/mm² respectively, and 90 directions per shell, along with eighteen reference (b = 0) volumes. Both DWI and T1w data were acquired at a spatical resolution of 1.25 mm isotropic. To prepare the training data, we use DMRITool\cite{cheng2017single} to select six uniformly distributed diffusion encoding directions per shell. Model fitting was primarily conducted using DIPY. Ground truth DTI metrics were obtained through tensor fitting of fully sampled diffusion data using a weighted least squares method as the reference gold standard. Ground truth NODDI metrics were derived using the AMICO\cite{daducci2015accelerated} algorithm, also serving as gold standard references.

\subsection{Experimental Settings}
Our method was evaluated both qualitatively and quantitatively, and compared with traditional model fitting(MF) algorithms and deep learning-based methods, specifically including q\_DL \cite{golkov2016q}, 2D-CNN \cite{gibbons2019simultaneous}, and MESC-SD \cite{ye2019deep}. q\_DL performed voxel-wise estimation using MLP without accounting for neighborhood information. 2D-CNN reshaped the raw data into slices and performed slice-wise estimation.  MESC-SD considered the diffusion signals in a $3\times 3\times 3$ image patch. We adopt MESC-SD as the backbone. To reduce memory consumption and training time, the inputs of our framework are
matrices with $4\times 4\times 4$ patches in the spatial direction and diffusion-weighted signals in the angular direction, and the outputs are maps with $4\times 4\times 4$ in the spatial direction and eleven diffusion parameters in the angular direction, and then average the results for overlapping voxels. Performance was evaluated using Peak Signal-to-Noise Ratio (PSNR) and Structural Similarity Index (SSIM).

\begin{figure}[!t]
  \centering
  \includegraphics[width=\linewidth]{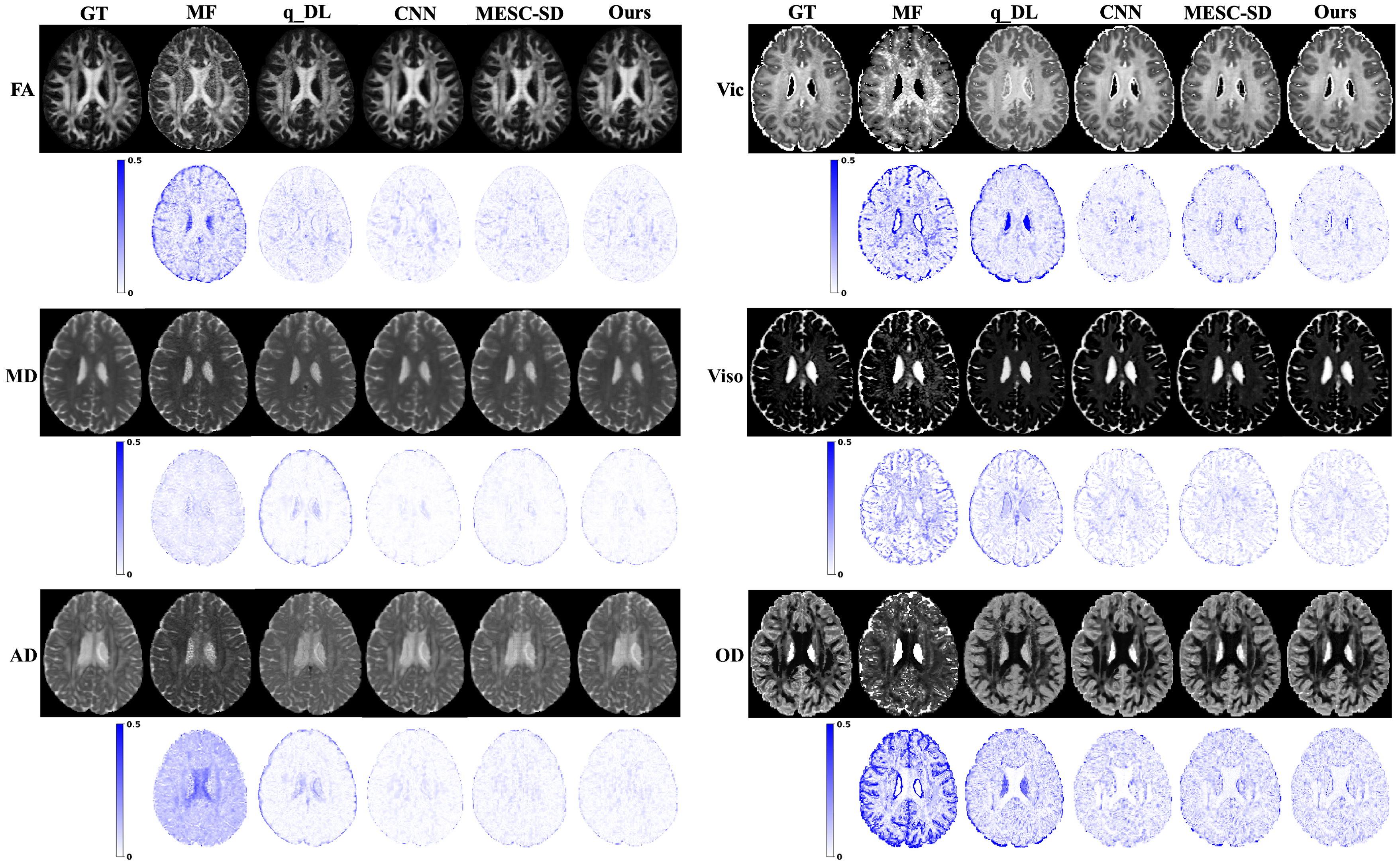} 
  \caption{The ground truth, estimated parametric maps and corresponding error maps based on MF, q\_DL, CNN, MESC-SD, and Ours in a test subject with 6 diffusion directions per shell at b-values of 1000, 2000, and 3000 $s/mm^{2} $.}
  \label{fig2}
\end{figure}

\section{Results}
\label{sec:format}
\subsection{Comparison of state-of-the-art methods}
Quantitative results are present in Table 1, demonstrating that the proposed method achieves better performance. Compared to the MESC-SD, the proposed method shows improvements over the baseline, validating its effectiveness in jointly estimating parametric maps. From the residual maps in Figure 2, it is evident that our method, incorporating the anatomical structure-guided module, successfully preserves fine-grained anatomical boundary details while effectively minimizing boundary noise. 

\subsection{Ablation Study: Comparison of Anatomical Priors}
To evaluate the effectiveness of the anatomical structure extraction module, we compared two types of priors: the structural image prior and the anatomical structure prior. The results, presented in Table 2, demonstrate that both priors perform well, further underscoring the importance of anatomical data in guiding the extraction of DW signal features. Furthermore, the results in Table 2 indicate that the method equipped with the anatomical structure module outperforms its counterparts, further confirming the module's effectiveness in filtering out anatomical boundary noise.

\begin{table}[!hpt]
\begin{center}
\caption{The quantitative results were obtained with 6 diffusion directions per shell at b-values of 1000, 2000, and 3000 $s/mm^{2} $. The best results are in bold.}
\label{tab1}
\resizebox{0.48\textwidth}{!}{
\begin{tabular}{c|lllllll}
\hline
Methods                        & \multicolumn{7}{c}{PSNR}                                                                                                                                                         \\ \cline{2-8} 
\multicolumn{1}{l|}{(18 DWIs)} & \multicolumn{1}{c}{FA} & \multicolumn{1}{c}{MD} & \multicolumn{1}{c}{AD} & \multicolumn{1}{c}{Vic} & \multicolumn{1}{c}{Viso} & \multicolumn{1}{c}{OD} & \multicolumn{1}{c}{Average} \\ \hline
MF                             & 22.5003                & 28.2471                & 22.5526                & 18.6438                 & 25.9331                  & 18.6266                & 21.5238±0.5762          \\ \hline
q\_DL                          & 29.9900                & 29.8063                & 29.0636                & 21.3592                 & 26.6373                  & 22.9165                & 25.2923±0.5505          \\ \hline
CNN                            & 33.8065              & 31.2779               & 30.6627          & 23.6787                & 29.3999                   & 24.6650               & 27.4668±0.6532         \\ \hline
MESC-SD             & 33.5589                & 31.5407                & 30.8925        & 24.1782           & 29.4174                & 25.1341               & 27.8358±0.6550          \\ \hline
\textbf{Ours}                  & \textbf{34.4223}       & \textbf{32.1819}       & \textbf{31.5219}       & \textbf{28.7883}        & \textbf{31.3110}         & \textbf{27.8336}       & \textbf{30.5126±0.5758} \\ \hline
Methods                        & \multicolumn{7}{c}{SSIM}                                                                                                                                                         \\ \cline{2-8} 
(18 DWIs)                      & \multicolumn{1}{c}{FA} & \multicolumn{1}{c}{MD} & \multicolumn{1}{c}{AD} & \multicolumn{1}{c}{Vic} & \multicolumn{1}{c}{Viso} & \multicolumn{1}{c}{OD} & \multicolumn{1}{c}{Average} \\ \hline
MF                             & 0.8686                 & 0.9577                 & 0.9183                 & 0.8735                  & 0.9369                   & 0.8492                 & 0.9007±0.0114           \\ \hline
q\_DL                          & 0.9418                 & 0.9594                 & 0.9471                 & 0.9160                  & 0.9336                   & 0.9143                 & 0.9354±0.0084           \\ \hline
CNN                            & 0.9664                & 0.9742                 & 0.9674         & 0.9575                   & 0.9598                   & 0.9474                 & 0.9621±0.0052           \\ \hline
MESC-SD        & 0.9647                 & 0.9728                 & 0.9657                 & 0.9563                  & 0.9598           & 0.9494             & 0.9615±0.0053           \\ \hline
\textbf{Ours}                  & \textbf{0.9712}        & \textbf{0.9753}        & \textbf{0.9690}        & \textbf{0.9776}         & \textbf{0.9681}          & \textbf{0.9680}        & \textbf{0.9715±0.0041}  \\ \hline

\end{tabular}}
\end{center}
\end{table}

\begin{table}[!hpt]
\begin{center}
\caption{The quantitative comparison of PSNR and SSIM with different anatomical priors. The results were obtained with 6 diffusion directions per shell at b-values of 1000, 2000, and 3000 $s/mm^{2} $}
\setlength{\intextsep}{-5pt} 
\label{tab2}
\resizebox{0.48\textwidth}{!}{
\begin{tabular}{c|ccccccc}
\hline
Methods         & \multicolumn{7}{c}{PSNR}                                                                                                                  \\ \cline{2-8} 
(18 DWIs)       & FA               & MD               & AD               & Vic              & Viso             & OD               & Average                    \\ \hline
DWIs            & 33.5589                & 31.5407                & 30.8925        & 24.1782           & 29.4174                & 25.1341               & 27.8358±0.6550          \\ \hline
DWIs+T1w        & 34.0326          & \textbf{32.3151} & \textbf{31.6293} & 28.3280          & 31.0744          & 27.5306          & 30.2753±0.6023          \\ \hline
DWIs+T1w tissue & \textbf{34.4223} & 32.1813          & 31.5219          & \textbf{28.7883} & \textbf{31.3110} & \textbf{27.8336} & \textbf{30.5126±0.5758} \\ \hline
Methods         & \multicolumn{7}{c}{SSIM}                                                                                                                  \\ \cline{2-8} 
(18 DWIs)       & FA               & MD               & AD               & Vic              & Viso             & OD               & Average                     \\ \hline
DWIs           & 0.9647                 & 0.9728                 & 0.9657                 & 0.9563                  & 0.9598           & 0.9494             & 0.9615±0.0053         \\ \hline
DWIs+T1w        & 0.9594           & 0.9731           & 0.9679           & 0.9755           & 0.9669           & 0.9657           & 0.9681±0.0136           \\ \hline
DWIs+T1w tissue & \textbf{0.9712}  & \textbf{0.9753}  & \textbf{0.9690}  & \textbf{0.9776}  & \textbf{0.9681}  & \textbf{0.9680}  & \textbf{0.9715±0.0041}  \\ \hline

\end{tabular}}
\end{center}
\end{table}

\section{CONCLUSION}
In this study, we propose an extendable framework for simultaneous estimation of parametric maps from multiple diffusion models using sparsely sampled q-space data. The proposed anatomical structure extraction module effectively guides DW signal feature extraction, filters out anatomical boundary noise, and achieves high-fidelity estimates with clearer tissue details. This flexible joint estimation framework leverages shared anatomical and diffusion information across multiple diffusion models to enhance estimation performance. Experimental results demonstrate that our method outperforms four state-of-the-art methods in both qualitative and quantitative evaluations. In future work, we aim to extend our approach to 5.0T brain imaging dataset \cite{wang2024diff5t} and a wider range of diffusion models.





\vspace{-5pt}
\section{Compliance with ethical standards}
\label{sec:ethics}
This research study was conducted retrospectively using human subject data made available in open access by HCP\cite{van2013wu}. Ethical approval was not required as confirmed by the license attached with the open access data.

\vspace{-5pt}
\section{Acknowledgments}
\label{sec:acknowledgments}

This research was partly supported by Shenzhen Medical Research Fund (No.B2402047), the National Natural Science Foundation of China (No.62222118, No.U22A2040), Shenzhen Science and Technology Program (No.RCYX20210706092104034, No.JCYJ20220531100213029), Key Laboratory for Magnetic Resonance and Multimodality Imaging of Guangdong Province (No.2023-B1212060052), and Youth lnnovation Promotion Association CAS.

\bibliographystyle{IEEEbib}
\bibliography{strings,main}  

\end{document}